\documentclass[a4paper, amsfonts, amssymb, amsmath, reprint, showkeys, nofootinbib, twoside, aps, pra, longbibliography]{revtex4-2}

\usepackage[T1]{fontenc}
\usepackage{graphicx} 
\newcommand\blfootnote[1]{%
	\begingroup
	\renewcommand\thefootnote{}\footnote{#1}%
	\addtocounter{footnote}{-1}%
	\endgroup
}

\usepackage{hyperref}
\hypersetup{
    plainpages=false,       
    unicode=false,          
    pdftoolbar=true,        
    pdfmenubar=true,        
    pdffitwindow=false,     
    pdfstartview={FitH},    
    pdfnewwindow=true,      
    colorlinks=true,        
    linkcolor=blue,         
    citecolor=blue,       
    filecolor=magenta,     
    urlcolor=cyan
}

\begin{document}

\title{Software compensation of trigger-synchronous control-frame errors in qubits and qudits}

\author{Gaurav A. Tathed$^{1,2,\dagger}$}
\author{Nicholas C.F. Zutt$^{1,2,\dagger}$}
\author{Collin J.C. Epstein$^{1,2}$}
\author{Crystal Senko$^{1,2,*}$}

\affiliation{$^1$Institute for Quantum Computing, University of Waterloo, Waterloo, Ontario, N2L 3G1, Canada}
\affiliation{$^2$Department of Physics and Astronomy, University of Waterloo, Waterloo, Ontario, N2L 3G1, Canada}

\blfootnote{$^\dagger$ These authors contributed equally.}
\blfootnote{$^*$ Corresponding author: csenko@uwaterloo.ca.}

\begin{abstract}

Quantum control experiments are often subject to coherent, time-dependent disturbances that vary over timescales comparable to the experiment duration. We show that when such disturbances are reproducible with respect to a trigger signal, their effect can be measured and compensated through software-defined updates to the control frequency and phase. We experimentally verify the performance of our protocol using a trapped $^{137}$Ba$^+$ ion experiencing magnetic-field-induced energy shifts synchronous with the laboratory AC mains power. Using this compensation technique, the calibrated AC line contribution to the instantaneous oscillator detuning is reduced by $21(9)\times$, and the fitted AC-induced phase amplitude is reduced below the measurement uncertainty. We use randomized benchmarking to validate the compensation performance in quantum gate sequences, recovering an average single-qubit gate fidelity of 99.93(1)\% with a magnetic-field-sensitive qubit. Finally, we extend the compensation framework to multi-level qudit control. Using the Bernstein-Vazirani algorithm as a benchmark, we increase the algorithm's success probability from 10(7)\% to 70(9)\% in a 16 level system when compensation is enabled. Our results demonstrate that trigger-synchronized coherent errors can be reframed as deterministic control-frame errors and corrected in software.

\end{abstract}

\maketitle

\section{Introduction}
\label{sec:1_Introduction}

Quantum computing platforms have advanced rapidly, with steady improvements in coherence, gate fidelity, control hardware, and system size \cite{wang21_singl_ion_qubit_with_estim, hughes25_trapp_ion_two_qubit_gates_with, ransford25_helios}.
At the same time, the field has developed a broad set of tools for diagnosing performance, including coherence measurements \cite{wang21_singl_ion_qubit_with_estim, mishin25_coher_microw_optic_qubit_level_neutr_thulium}, randomized benchmarking \cite{chen25_random_bench_with_leakag_error, sannamoth25_easier_random_gates_provid_more, gu23_bench_univer_quant_gates_via_chann_spect}, and algorithmic demonstrations \cite{wright19_bench_quant_comput, ringbauer22_univer_qudit_quant_proces_with_trapp_ions}.
As systems scale, maintaining reliable performance increasingly requires identifying and suppressing environmental noise that couples into the quantum device.

Many error models emphasize stochastic noise sources such as broadband dephasing, amplitude noise, and relaxation \cite{devitt13_quant_error_correc_begin, brown04_arbit_accur_compos_pulse_sequen}.
In practice, quantum experiments can also be affected by structured, coherent disturbances that leave repeatable signatures in time-domain measurements and benchmarking data \cite{hu23_compen_power_line_induc_magnet, wei22_measur_suppr_magnet_field_noise, laraoui10_magnet_random_ac_magnet_field}.
Unlike broadband stochastic noise, these disturbances are often time dependent and arise from reproducible environmental or technical sources, including vibrations, electrical pickup, power-supply ripple, and modulation-chain imperfections \cite{merkel19_magnet_field_stabil_system_atomic_physic_exper, you26_charac_activ_cancel_power_line, pagano19_cryog_trapp_ion_system_large, kono24_mechan_induc_correl_error_super, kalra16_vibrat_induc_elect_noise_cryog, kellermann24_vibrat_decoup_system_tes_operat}.
A prominent example is interference from the AC mains line signal at 50/60 Hz and its harmonics \cite{ruster16_long_lived_zeeman_trapp_ion_qubit, rowe02_trans_quant_states_separ_ions, haffner08_quant_comput_with_trapp_ions}.
Such disturbances can produce coherent errors by modulating the qubit or qudit transition frequency, drive amplitude, or optical phase through mechanisms such as Zeeman shifts, AC Stark shifts, oscillator detuning, laser-frequency excursions, or line-synchronous effects in the electronic modulation chain \cite{haffner08_quant_comput_with_trapp_ions, bruzewicz19_trapp_ion_quant_comput, you26_charac_activ_cancel_power_line}.

Existing mitigation strategies often target the physical coupling pathways.
These include passive shielding and pickup reduction \cite{ruster16_long_lived_zeeman_trapp_ion_qubit, kranzl22_contr_long_ion_strin_quant, brandl16_cryog_setup_trapp_ion_quant_comput}, active field cancellation using compensation coils or feedback \cite{kranzl22_contr_long_ion_strin_quant, hu23_compen_power_line_induc_magnet}, and broader stabilization of ambient magnetic-field fluctuations \cite{merkel19_magnet_field_stabil_system_atomic_physic_exper}.
Mechanically coupled noise can be mitigated through direct vibration isolation and dampening or through active monitoring and feedback of interferometer signals \cite{kalra16_vibrat_induc_elect_noise_cryog, pagano19_cryog_trapp_ion_system_large, kono24_mechan_induc_correl_error_super, kellermann24_vibrat_decoup_system_tes_operat}. 
More general closed-loop drift-control techniques can also improve stability, but typically require additional sensing, calibration, or monitoring signals \cite{vepsaelaeinen22_improv_qubit_coher_using_closed_loop_feedb, proctor20_detec_track_drift_quant_infor_proces}.
Although these approaches can be effective, they often require additional sensors, coils, shielding, calibration, or changes to the experimental apparatus, and their performance can be limited by bandwidth, latency, added noise, and spatial or channel-dependent coupling 
\cite{kranzl22_contr_long_ion_strin_quant}.
However, many structured disturbances are deterministic and reproducible when viewed in the trigger-referenced frame.
This reproducibility motivates a complementary software-based approach: rather than suppressing the disturbance at its source, its coherent effect can be measured, predicted, and compensated directly in the programmed control waveform.
Related phase-tracking ideas have been used to correct transport-induced phase accumulation in trapped-ion QCCD devices, where ions experience position- and time-dependent magnetic fields during shuttling \cite{arkin22_quant_comput_phase_track, moses23_race_track_trapp_ion_quant_proces, figgatt18_build_program_univ_trap_quant, kielpinski02_archit_large_scale_ion_trap_quant_comput}.

In this work, we present a framework for compensating trigger-synchronous detuning and dephasing errors through software-defined updates to the programmed control waveform.
We study the example of magnetic-field fluctuations synchronous with the AC mains line signal, which produce deterministic frequency shifts and accumulated phase errors when experiments are referenced to the line trigger.
Rather than suppressing the disturbance at its source, we measure its effect in the trigger-referenced frame and compensate the resulting detuning and phase evolution directly in software.
Experimentally, we demonstrate this approach at three levels.
First, using Ramsey-type measurements, we suppress both the instantaneous detuning errors and the resulting phase-accumulation errors.
Second, we test the compensation at the gate-sequence level using Haar-randomized benchmarking.
Finally, we extend the framework beyond two-level qubit control by implementing Bernstein--Vazirani algorithm circuits in a single trapped-ion qudit, showing that the same compensation principle applies naturally to qudits and more general multilevel quantum systems.

\begin{figure*}[ht!] 
  \includegraphics[width=1\textwidth]{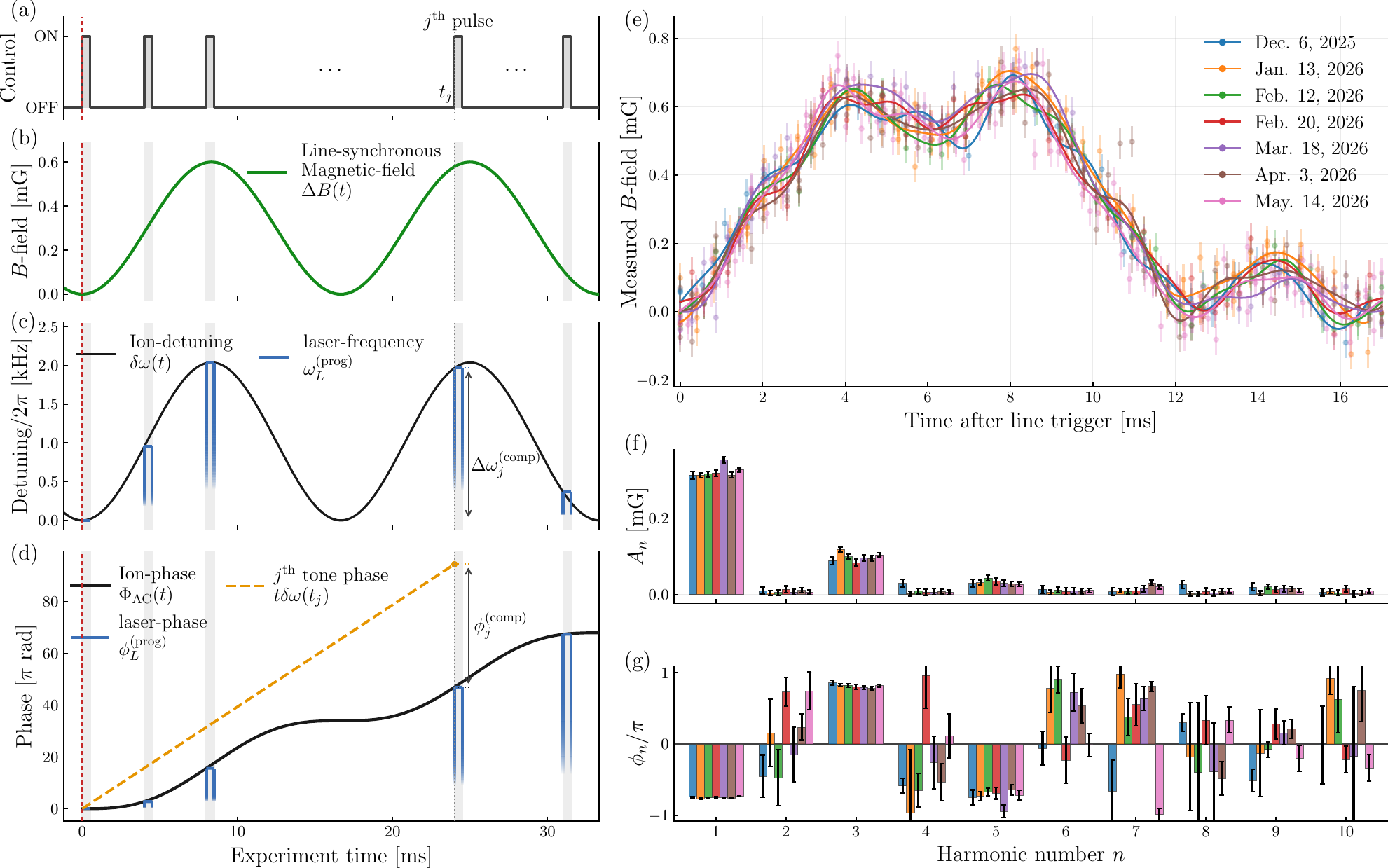}
\caption{
Compensation protocol and AC line-synchronous magnetic-field reproducibility. 
(a) Control sequence starting at the line-trigger time origin. 
All pulse frequencies and phases are programmed relative to a common waveform start, with pulse $j$ applied at start time $t_j$. 
(b) Example line-synchronous magnetic-field perturbation $\Delta B(t)$ (at purely 60~Hz) in the line-trigger-referenced time coordinate.
(c) Corresponding transition frequency detuning and programmed laser frequency. 
For pulse $j$, the applied frequency offset is chosen to match the instantaneous detuning, $\Delta\omega_j^{(\mathrm{comp})}=\delta\omega(t_j)$, as indicated by the vertical arrow. 
(d) Phase accumulated by the qubit, $\Phi_{\mathrm{AC}}(t)$, due to the line-synchronous detuning, together with the programmed laser phase during the pulse windows, in the frame rotating at the reference frequency $\omega_0$. 
The dashed orange line shows the phase, $t \delta\omega(t_j),$ accumulated by the constant-frequency tone used for pulse $j$ when referenced to the global start time $t=0$. 
The vertical arrow indicates the phase correction $\phi_j^{(\mathrm{comp})}$ required to match the globally referenced oscillator phase to the qubit's phase. 
(e) Repeated measurements of the line-synchronous magnetic-field perturbation over multiple dates, showing that the waveform is reproducible in the line-trigger-referenced frame. 
(f) Fitted harmonic amplitudes $A_n$ of the measured waveforms in (e) for the first ten harmonics of the line frequency. 
For the dominant $60$, $180$, and $300~\mathrm{Hz}$ components, the across-date mean amplitudes and standard deviations are 
$0.321(15)$, $0.097(11)$, and $0.032(6)~\mathrm{mG}$, respectively. 
(g) Corresponding fitted harmonic phases $\phi_n/\pi$. 
For the same $60$, $180$, and $300~\mathrm{Hz}$ components, the circular mean phases are 
$-0.749\pi$, $0.813\pi$, and $-0.735\pi$, with circular standard deviations 
$0.011\pi$, $0.024\pi$, and $0.091\pi$, respectively. 
The reproducibility of both the time-domain waveform and its harmonic components enables deterministic software compensation of the line-synchronous perturbation.
}
    \label{fig:compensation_schematic}
\end{figure*}

\section{Compensation protocol}
\label{sec:compensation_scheme}

We consider a line-synchronous perturbation that changes the detuning between an addressed qudit transition frequency and the programmed control tone. 
Because the experiment is synchronized to the line trigger, the reproducible component of this perturbation can be written as a deterministic time-dependent detuning $\delta\omega(t)$ in a line-trigger-referenced time coordinate. 
In the experiments below, we see that the dominant contribution couples as a reproducible time-dependent magnetic-field fluctuation, but the compensation protocol is equally applicable no matter the physical origin, requiring only knowledge of the resulting detuning waveform $\delta\omega(t)$.
The central idea of the protocol is that this deterministic detuning produces two predictable errors: an instantaneous frequency error during each pulse, and a phase error accumulated between pulses. 
Both can be corrected in control software by updating the frequency and phase of the programmed oscillator tone that drives transitions.

Let $\omega_0$ denote the reference transition frequency. 
The applied control field is taken to be proportional to
\begin{equation}
\Omega(t)\cos\!\big(\theta_{\mathrm{L}}(t)\big),
\qquad
\theta_{\mathrm{L}}(t)=\omega_{\mathrm{L}}t+\phi_{\mathrm{L}}(t),
\end{equation}
where $\Omega(t)$ is the drive amplitude, $\omega_{\mathrm{L}}$ is the nominal oscillator frequency, and $\phi_{\mathrm{L}}(t)$ is a programmable phase. 

Fig. \ref{fig:compensation_schematic}(a-d) illustrates the compensation protocol for a pulse sequence whose phases are referenced to a common $t=0$ (start of the experiment). 
Let pulse $j$ begin at time $t_j$. 
The atomic transition frequency at that time is
\begin{equation}
\label{eq:freq_offset_impl}
\omega_a(t_j)
=
\omega_0+\delta\omega(t_j),
\end{equation}
where $\omega_0$ is the chosen reference transition frequency, defined by the atomic resonance at $t=0$, and $\delta\omega(t)$ is the time-dependent shift relative to that reference.

If the line-synchronous perturbation, and therefore the transition frequency, varies negligibly over the duration of pulse $j$, then the pulse can be made resonant by choosing the programmed oscillator frequency to match the instantaneous transition frequency at $t_j$:
\begin{equation}
\omega_{L,j}^{(\mathrm{prog})}
=
\omega_0+\Delta\omega_j^{(\mathrm{comp})}
=
\omega_0+\delta\omega(t_j)
=
\omega_a(t_j).
\label{eq:detuning_comp_impl}
\end{equation}
Here $\Delta\omega_j^{(\mathrm{comp})}$ is the applied frequency correction relative to the reference frequency $\omega_0$. 
Thus, detuning compensation corresponds to choosing
$\Delta\omega_j^{(\mathrm{comp})}=\delta\omega(t_j)$, so that the programmed oscillator frequency equals the instantaneous transition frequency, as shown in Fig.~\ref{fig:compensation_schematic}c.

Ensuring each pulse is resonant does not by itself guarantee that the phase of pulse $j$ is consistent with the phase accumulated by the qudit states since the start of the sequence. 
A separate phase update is therefore required.
Taking the start of the sequence as the global phase reference, the relative phase accumulated by an atomic superposition due to the AC line-synchronous detuning up to time $t$ is
\begin{equation}
\Phi_{\mathrm{AC}}(t)
=
\int_0^t \delta\omega(t')\,dt'.
\label{eq:phase_due_to_line_signal}
\end{equation}
Thus, once $\delta\omega(t)$ is measured in the line-trigger-referenced time coordinate, the corresponding phase evolution can be calculated. 
A rotating-frame derivation of Eq.~\eqref{eq:phase_due_to_line_signal} and its connection to the compensation condition is given in Methods~\ref{sec:compensation_derivation}.

An ideal compensation scheme would make the control oscillator continuously track the instantaneous transition frequency. 
In that limit, both the frequency and phase of the oscillator would follow the line-synchronous perturbation at all times. 
In the lab frame, the tracked oscillator phase is
\begin{equation}
\theta_{\mathrm{track}}(t)
=
\theta_\text{L}(0)+\omega_0 t+\Phi_{\mathrm{AC}}(t),
\label{eq:theta_track_impl}
\end{equation}
where $\theta_\text{L}(0)$ is the oscillator phase at $t=0$. 

In practice, continuously updating the oscillator frequency throughout the entire sequence is unnecessary. 
Because each pulse is short compared with the timescale over which $\delta\omega(t)$ varies, we approximate the tracked waveform by updating the oscillator frequency and phase only at the start of the pulse. 
The role of the phase update is therefore to make the phase of pulse $j$ agree with the ideal tracked phase in Eq.~\eqref{eq:theta_track_impl} at the time $t_j$ when the pulse is applied.
Now suppose pulse $j$ is implemented using the constant frequency $\omega_{L,j}^{\text{(prog)}}$ specified in Eq.~\eqref{eq:detuning_comp_impl}. 
Because this tone is phase-referenced to the experiment start time, $t=0$, its phase at time $t_j$ before adding the programmed pulse phase is
\begin{equation}
\theta_{\mathrm{inst}}(t_j)
=
\theta_L(0)+\big[\omega_0+\delta\omega(t_j)\big]t_j .
\label{eq:theta_inst_impl}
\end{equation}

The programmed phase is chosen so that the phase of this constant-frequency tone at the pulse time matches the ideal continuously tracked phase, while still implementing the desired logical pulse phase. 
Therefore,
\begin{equation}
\theta_{\mathrm{inst}}(t_j)+\phi_j^{(\mathrm{prog})}
=
\theta_{\mathrm{track}}(t_j)+\phi_j^{(\mathrm{ideal})}.
\end{equation}
We express the programmed oscillator phase $\phi_{L,j}^{(\mathrm{prog})}$ in terms of the logical pulse phase $\phi_j^{(\mathrm{ideal})}$ and the compensating phase $\phi_j^{(\mathrm{comp})}$ as
\begin{equation}
\phi_{L,j}^{(\mathrm{prog})}
=
\phi_j^{(\mathrm{ideal})}
+
\phi_j^{(\mathrm{comp})},
\end{equation}
which yields the expression for the phase compensation as
\begin{equation}
\begin{aligned}
\phi_j^{(\mathrm{comp})}
&=
\theta_{\mathrm{track}}(t_j)-\theta_{\mathrm{inst}}(t_j) \\
&=
\Phi_{\mathrm{AC}}(t_j)-t_j \delta\omega(t_j) \\
&=
\int_0^{t_j}\delta\omega(t')\,dt'
-
t_j\,\delta\omega(t_j).
\end{aligned}
\label{eq:phase_comp_impl}
\end{equation}

Equations~\eqref{eq:detuning_comp_impl} and \eqref{eq:phase_comp_impl} define the full pulse-level compensation rule:
\begin{align}
\omega_{L,j}^{(\mathrm{prog})} &= \omega_0+\delta\omega(t_j), \\
\phi_{L,j}^{(\mathrm{prog})} 
&= 
\phi_j^{(\mathrm{ideal})}
+
\int_0^{t_j}\delta\omega(t')\,dt'
-
t_j\,\delta\omega(t_j).
\end{align}
The first update makes pulse $j$ resonant at its application time, while the second ensures that its phase correctly tracks that of the qubit or transition being addressed.

Specializing to a line-synchronous magnetic-field perturbation small enough that the transition frequency shifts linearly with magnetic field, the detuning for pulse $j$ can be written as

\begin{figure*}[t] 
  \includegraphics[width=1\textwidth]{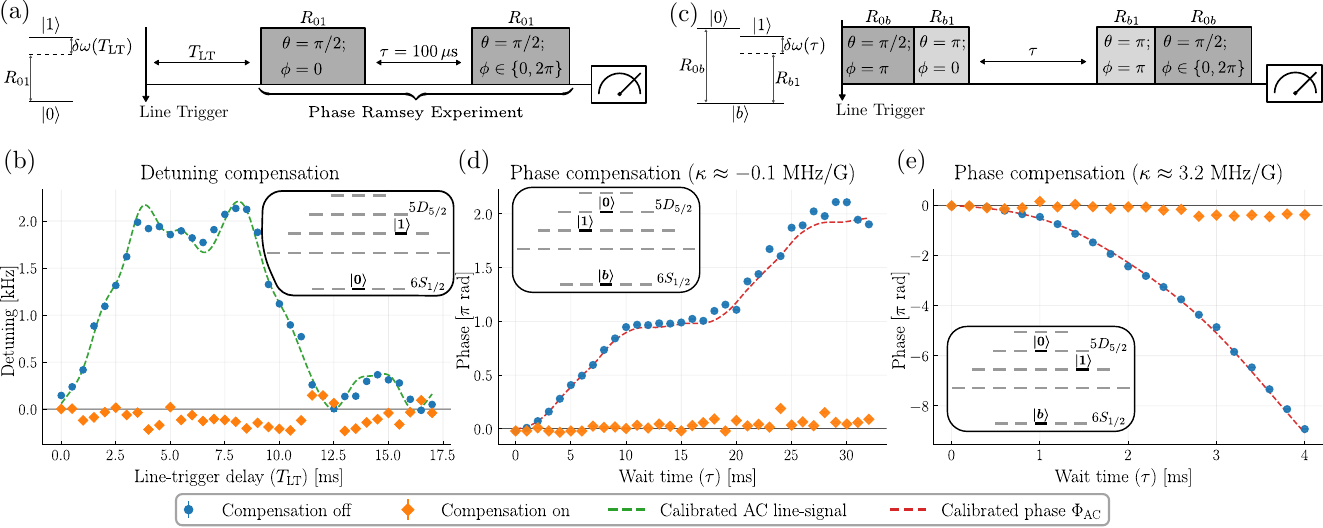}
    \caption{
    Experimental measurement and compensation of AC line-synchronous detuning and phase errors. 
    (a) Pulse sequence used to measure the instantaneous AC line-synchronous detuning. 
    Each shot is synchronized to the line trigger, followed by a variable trigger delay $T_{\mathrm{LT}}$, and then a fixed-time Ramsey experiment. 
    The analyzer phase is varied to extract the transition detuning at the selected trigger delay. 
    (b) Measured detuning versus $T_{\mathrm{LT}}$ with compensation disabled and enabled. 
    The trigger delay is scanned over one full period of the line signal to reconstruct the detuning throughout the AC cycle. 
    Without compensation, the fitted detuning follows the calibrated AC line signal, shown by the dashed green curve. 
    With compensation enabled, the residual detuning is strongly suppressed; the matched-filter scale is reduced by $21(9)\times$, and the fitted harmonic amplitude is reduced by $8(1)\times$; see Table~\ref{tab:compensation_suppression}. 
    (c) Pulse sequence used to measure accumulated AC phase. 
    A hyperfine qubit Ramsey experiment is performed after the line trigger, with variable wait time $\tau$, to resolve the phase accumulated between the Ramsey pulses. 
    The qubit is coupled optically through the bus state $|b\rangle=|6S_{1/2},F=2,m_F=0\rangle$. 
    The $|0\rangle\leftrightarrow |b\rangle$ optical transition is first-order magnetic-field insensitive, with sensitivity of only approximately $40~\mathrm{Hz/G}$, so the magnetic-field sensitivity of the hyperfine qubit is dominated by the $|1\rangle$ state. 
    (d) Measured phase residual for a weakly magnetic-field-sensitive hyperfine qubit, $\kappa\approx -0.1~\mathrm{MHz/G}$. 
    The wait time $\tau$ is scanned over approximately two periods of the line signal. 
    Without compensation, the phase accumulates as expected from the AC line-synchronous  detuning, $\Phi_{\mathrm{AC}}$, shown by the dashed red curve.
    With phase compensation enabled, the fitted AC phase amplitude is reduced below the measurement uncertainty; the central-value matched-filter reduction is $90\times$, and the fitted harmonic amplitude is reduced by $4.7(1.1)\times$; see Table~\ref{tab:compensation_suppression}. 
    (e)~Corresponding phase-compensation measurement for a more magnetic-field-sensitive qubit, $\kappa\approx 3.2~\mathrm{MHz/G}$. 
    Because the larger magnetic-field sensitivity leads to faster dephasing, this measurement is restricted to a shorter wait time. 
    The larger uncompensated phase accumulation is again described by the calibrated AC phase, and compensation keeps the residual phase near zero.
    In panels (b),(d) and (e) the error bars (fit uncertainty in the extracted detuning or phase) are smaller than the marker size.
    }
    \label{fig:detuning_phase_compensation}
\end{figure*}

\begin{equation}
\delta\omega_j(t)=2\pi\kappa_j \Delta B(t),
\end{equation}
where $\kappa_j$ is the magnetic-field sensitivity of the transition addressed by pulse $j$. 
The compensation rules then become
\begin{align}
\omega_{L,j}^{(\mathrm{prog})} &= \omega_0+2\pi \kappa_j \Delta B(t_j), \\
\phi_{L,j}^{(\mathrm{prog})}
&= \phi_j^{(\text{ideal})} + 
2\pi \kappa_j
\left[
\int_0^{t_j}\Delta B(t')\,dt'
-
t_j \Delta B(t_j)
\right].
\label{eq:phase_comp_B_impl}
\end{align}
Any static (DC) component of $B(t)$ cancels in Eq.~\eqref{eq:phase_comp_B_impl}, so the phase correction depends only on the time-varying part of the perturbation. This compensation framework can also be extended to multilevel systems, see Methods~\ref{sec:extension_to_qudits}.
The protocol assumes that the measured trigger-synchronous waveform remains stable between calibration and the compensated experiment, and that the control system can update the programmed frequency and phase with sufficient timing resolution. 
Fig. \ref{fig:compensation_schematic}(e-g) shows the reproducibility of the measured AC line-synchronous magnetic field perturbations over several months. We see that for the dominant $60$, $180$, and $300~\mathrm{Hz}$ components, the across-date mean amplitudes (and relative standard deviations) are 
$0.321(4.6\%)$, $0.097(11\%)$, and $0.032(19\%)~\mathrm{mG}$, respectively. 

\section{Results}

We implement the compensation protocol using a single $^{137}\mathrm{Ba}^+$ ion confined in a linear Paul trap (see Methods~\ref{sec:experimental_setup} for experimental setup details). Coherent control is implemented with a 1762~nm laser locked to a high-finesse optical cavity \cite{bramman23_ablat_loadin_qudit_measur_barium}. 
This field drives the electric-quadrupole transition between hyperfine sublevels of the $6S_{1/2}$ and $5D_{5/2}$ manifolds. 
Different transitions are addressed by changing the modulation frequency applied to an electro-optic modulator. 
The radio-frequency tones used for this modulation are generated by an RFSoC-based control system, so the frequency and phase of each optical tone can be programmed independently on a pulse-by-pulse basis.

The results below quantify the performance of this software compensation protocol across several experimental settings, from direct measurements of line-synchronous detuning and phase accumulation to gate-sequence benchmarking and multilevel qudit circuits.

\subsection{Compensating the line-synchronous signal}

We first calibrate the line-synchronous perturbation in the same line-trigger-referenced time coordinate used for compensation. 
Because the perturbation is reproducible with respect to the line trigger, its effect can be reconstructed by sampling the transition frequency at different delays within the line cycle, see Methods~\ref{sec:measure_line_sig} for details. 
The resulting detuning waveform, shown in Fig. \ref{fig:compensation_schematic}(e-g), then provides the calibration input for both the frequency and phase corrections described above.

\begin{table*}[t]
\centering
\caption{
Quantification of detuning and phase compensation. 
The matched-filter scale $a_{\mathrm{AC}}$ gives the fitted amplitude of the calibrated AC waveform in the measured signal. 
The harmonic amplitude $A_{\mathrm{AC}}=\sqrt{\sum_n A_n^2}$ quantifies the total fitted amplitude at the line frequency and its harmonics. 
Suppression factors are computed from the ratio of the uncompensated and compensated values.
}
\label{tab:compensation_suppression}
\begin{tabular*}{\textwidth}{@{\extracolsep{\fill}}lccc}
\hline
Measurement & Compensation off & Compensation on & Suppression \\
\hline
\hline
\multicolumn{4}{l}{\textit{Matched-filter scale, $a_{\mathrm{AC}}$}} \\
Detuning & $0.98(2)$ & $-0.05(2)$ & $21(9)\times$ \\
Phase    & $1.1(1)$  & $0.01(5)$  & $^*90\times$ \\
\hline
\multicolumn{4}{l}{\textit{Harmonic amplitude, $A_{\mathrm{AC}}$}} \\
Detuning & $1.09(2)~\mathrm{kHz}$ & $0.13(1)~\mathrm{kHz}$ & $8(1)\times$ \\
Phase & $0.19(1)\pi~\mathrm{rad}$ & $0.041(9)\pi~\mathrm{rad}$ & $5(1)\times$ \\
\hline
\end{tabular*}
$^*$ The compensated phase projection is statistically consistent with zero; therefore we report the fitted amplitudes rather than assigning a precise uncertainty to the suppression ratio.
\end{table*}
We directly test the two components of the compensation protocol: the frequency update that makes each pulse resonant at its application time, and the phase update that cancels the accumulated line-synchronous phase. 
Both experiments use the same line-trigger reference used to calibrate the line-synchronous waveform, so that the compensation is evaluated at a known time within the line cycle.

For the detuning-compensation measurement, we use the Ramsey sequence shown in Fig.~\ref{fig:detuning_phase_compensation}(a). 
This experiment is similar to the sequence used to measure the line-synchronous perturbation, but here the goal is to compare the residual detuning with the compensation disabled and enabled. 
Each shot is synchronized to the line trigger, followed by a variable delay $T_{\mathrm{LT}}$, and then a Ramsey experiment with fixed free-evolution time $\tau=100~\mu\mathrm{s}$. 
The Ramsey analyzer phase is sampled at ten equally spaced values between $0$ and $2\pi$, and the resulting population oscillation is fit to extract the detuning.

The detuning-compensation experiment is performed on the relatively magnetic-field-sensitive optical qubit,
$|0\rangle=|{6S_{1/2},F=2,m_F=0}\rangle$, 
and $|{1}\rangle=|{5D_{5/2},F=3,m_{F}=2}\rangle$,
which has magnetic-field sensitivity $\kappa \approx 3.2~\mathrm{MHz/G}$. 
For each value of $T_{\mathrm{LT}}$, the detuning is extracted by fitting the measured Ramsey data to the same finite-pulse numerical model used in the line-signal calibration. 
As shown in Fig.~\ref{fig:detuning_phase_compensation}(b), with compensation disabled the fitted detuning varies strongly over the line cycle. 
Using the matched-filter analysis described in Methods~\ref{sec:compensation_quantification}, the fitted calibrated AC component is reduced from $a_{\mathrm{AC}}=0.98(2)$ without compensation to $a_{\mathrm{AC}}=-0.048(2)$ with compensation, corresponding to a $20.6\times$ suppression of the calibrated AC waveform. 
The total harmonic amplitude is reduced from $A_{\mathrm{AC}}=1.09~\mathrm{kHz}$ to $0.132~\mathrm{kHz}$, giving an independent $8.32\times$ suppression of residual line-frequency harmonic content; these values are summarized in Table~\ref{tab:compensation_suppression}. 
Thus, when the programmed oscillator frequency is updated according to the measured line-synchronous waveform, the residual detuning is strongly suppressed, showing that the control tone tracks the instantaneous transition frequency.

We then test the phase-compensation component using the sequence shown in Fig.~\ref{fig:detuning_phase_compensation}(c). 
For this measurement, we use a hyperfine qubit rather than a directly coupled optical qubit. 
For an optical qubit, the Ramsey coherence in our system is limited by 1762~nm laser frequency noise; we measure a coherence time of approximately $T_2^*=1.5$~ms. 
This is much shorter than the period of the line-synchronous perturbation, so optical Ramsey experiments over comparable timescales are dominated by laser phase noise rather than the magnetic-field-induced phase we wish to isolate. 
We therefore encode the qubit in a pair of hyperfine states whose control sequence is arranged so that optical phase fluctuations are largely common mode, while sensitivity to the line-synchronous magnetic-field perturbation is retained.
We use a hyperfine qubit encoded within the $5D_{5/2}$ manifold,
$|0\rangle=|5D_{5/2},F=2,m_F=0\rangle$ and
$|1\rangle=|5D_{5/2},F=3,m_F=-1\rangle$.
The two qubit states are coupled through the bus state $|b\rangle=|6S_{1/2},F=2,m_F=0\rangle$,
using the optical control field. 
This encoding suppresses sensitivity to common optical phase noise while retaining sensitivity to the line-synchronous magnetic-field perturbation. 
The magnetic-field sensitivity of this hyperfine qubit is approximately $\kappa \simeq -0.1~\mathrm{MHz/G}$.
To measure the accumulated phase, we vary the wait time $\tau$ relative to the line trigger and extract the Ramsey phase. 
Without compensation, the measured phase residual follows the line-synchronous phase accumulated between the Ramsey pulses given by Eq.~\eqref{eq:phase_due_to_line_signal}. 
The matched-filter scale is $a_{\mathrm{AC}}=1.1(1)$ without compensation and is reduced to $a_{\mathrm{AC}}=0.01(5)$ with compensation. 
Because the compensated value is consistent with zero, the suppression ratio is not well conditioned; the ratio of the best-fit values corresponds to a $90\times$ reduction, but the primary conclusion is that the residual calibrated AC phase amplitude is below the measurement uncertainty. 
The harmonic amplitude is reduced from $0.19(1)\pi~\mathrm{rad}$ to $0.041(9)\pi~\mathrm{rad}$, corresponding to $\sim 5(1)\times$ suppression of residual line-frequency harmonic content, as summarized in Table~\ref{tab:compensation_suppression}. 
With phase compensation enabled, the programmed pulse phases are updated using the calibrated waveform, and the residual phase remains near zero, as shown in Fig.~\ref{fig:detuning_phase_compensation}(d). 
To further test phase compensation at larger magnetic-field sensitivity, we repeat the accumulated-phase measurement using a hyperfine qubit with approximately the same sensitivity, $\kappa \approx 3.2~\mathrm{MHz/G}$, as the transition used in the detuning-compensation measurement of Fig. 2(b). Because this larger sensitivity leads to faster dephasing and a coherence time shorter than the period of the line-synchronous signal, the phase accumulation cannot be measured over a full line cycle; nevertheless, Fig. 2(e) shows that compensation keeps the residual phase near zero over the accessible time window.

Together, these measurements show that the same line-synchronous calibration can be used to correct both the instantaneous detuning during driven pulses and the phase accumulated between pulses, across transitions with different magnetic-field sensitivities.

\subsection{Benchmarking gate-sequences under line-synchronous noise}

To demonstrate that our compensation protocol decreases the error of quantum gate sequences, we utilize a randomized benchmarking technique.
Randomized benchmarking is a standard method for quantifying the performance of quantum gates while reducing sensitivity to state-preparation and measurement errors.
Instead of characterizing each gate individually, the experiment applies sequences of randomly chosen gates, followed by a final recovery operation that ideally returns the system to its initial state. 
As the sequence length increases, gate errors accumulate and the probability of recovering the initial state decays.
Fitting this decay provides a compact measure of the average error per gate \cite{emerson05_scalab_noise_estim_with_random_unitar_operat, knill08_random_bench_quant_gates, magesan12_effic_measur_quant_gate_error}.
The usual decay model is given by 
\begin{equation}
\label{eq:rb_decay}
P(m)=A p^m+B,
\end{equation}
where $m$ is the number of Haar-sampled random unitaries, $p$ is the fitted decay parameter, and $A$ and $B$ absorb state-preparation and measurement errors.
For a single qubit, the reported average fidelity is obtained from the fitted depolarizing parameter as
\begin{equation}
F_{\mathrm{avg}}=\frac{1+p}{2}.
\end{equation}
This provides a compact, effective summary of the gate performance, but it does not explicitly capture sequence-dependent deviations caused by deterministic line-synchronous noise.

This decay model assumes an effectively stationary, Markovian error channel during the randomized sequences. Line-synchronous noise violates this assumption because the detuning depends on the gate's timing within the trigger-referenced line cycle, introducing time-correlated errors and deviations from the simple exponential decay.
We therefore use the fitted fidelity as an operational benchmark, while interpreting deviations from the simple exponential decay as a signature of correlated, trigger-referenced detuning errors.
Higher-fidelity under compensation, and a better fit to the decay model, therefore signify the removal of this time-correlated error channel.

We test this effect by performing randomized benchmarking on two optical qubits with different magnetic-field sensitivities. 
Both qubits share the same ground state,
$|0\rangle \equiv |{6S_{1/2},F=2,m_F=0}\rangle$, but use different excited states in the $5D_{5/2}$ manifold. 
The magnetic-field-sensitive qubit is encoded with 
$|{1}\rangle\equiv |{5D_{5/2},F=3,m_F=2}\rangle$, giving a transition sensitivity of $\sim 3.2~\mathrm{MHz/G}$. 
The magnetic-field-insensitive qubit is encoded with 
$|{1}\rangle\equiv |{5D_{5/2},F=2,m_F=0}\rangle$, with a much smaller sensitivity of $\sim 40~\mathrm{Hz/G}$. 
The insensitive qubit therefore serves as a reference that is largely unaffected by the line-synchronous magnetic-field perturbation.

\begin{figure*}[t] 
  \includegraphics[width=0.97\textwidth]{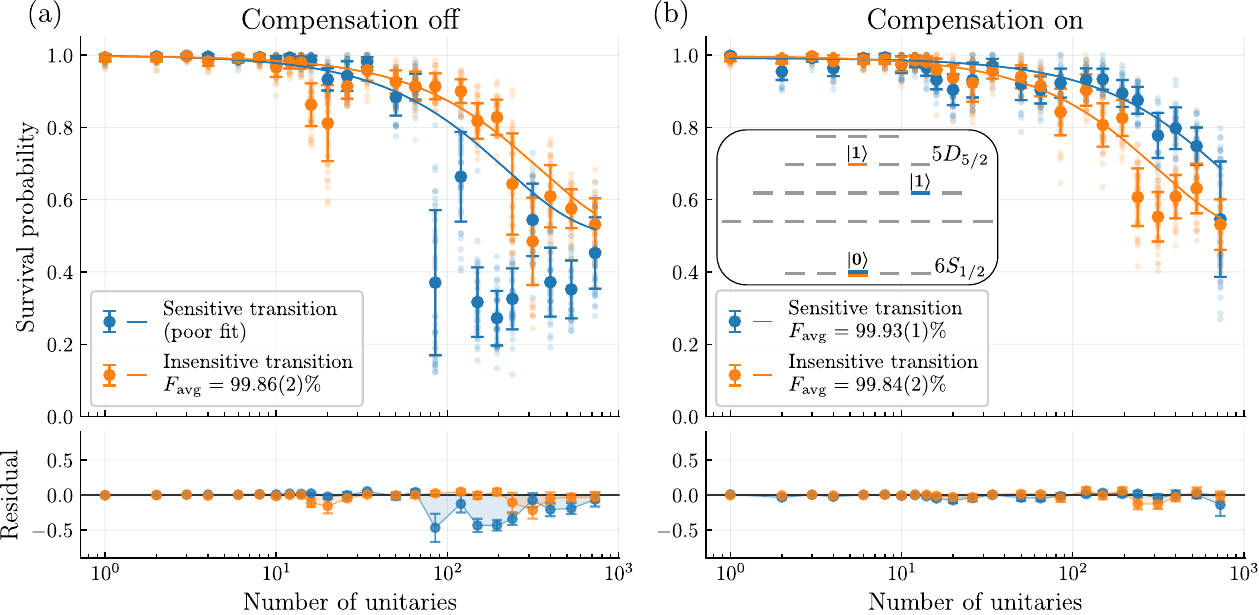}
    \caption{
    Haar-randomized benchmarking of magnetic-field-sensitive and magnetic-field-insensitive optical transitions with and without AC line-synchronous compensation. 
    (a) Randomized benchmarking with compensation disabled. 
    The magnetic-field-sensitive transition, with sensitivity $\kappa\approx 3.2~\mathrm{MHz/G}$, exhibits large sequence-dependent fluctuations and cannot be well described by the zeroth-order randomized-benchmarking decay model, indicating coherent errors from uncompensated line-synchronous detuning during the random sequences. 
    The comparatively magnetic-field-insensitive transition, with sensitivity approximately $40~\mathrm{Hz/G}$, is less affected and gives a fitted average fidelity of $F_{\mathrm{avg}}=99.86(2)\%$. 
    (b) Randomized benchmarking with compensation enabled. 
    The sensitive transition recovers a smooth decay and improves to $F_{\mathrm{avg}}=99.93(1)\%$, comparable to (and even better than) the insensitive transition, which gives $F_{\mathrm{avg}}=99.84(2)\%$. 
    The inset level diagram shows the $6S_{1/2}$ and $5D_{5/2}$ states used for the magnetic-field-sensitive and magnetic-field-insensitive benchmarking experiments. 
    Faint points show individual random-sequence outcomes, filled markers with error bars show the mean and spread at each sequence length, and solid curves are fits to the zeroth-order randomized-benchmarking model $P(m)=Ap^m+B$. 
    Lower panels show the residuals of the fitted decay curves.
    }
    \label{fig:rb_compensation}
\end{figure*}

For each qubit, we perform benchmarking with the compensation disabled and enabled, while synchronizing each experimental shot to the same line-trigger reference. 
At each sequence length, the survival probability is averaged over 40 sets of unitaries sampled randomly according to the Haar measure, with each set measured over 100 experimental shots.

As shown in Fig.~\ref{fig:rb_compensation}(a), without compensation the magnetic-field-insensitive qubit (orange) yields a fitted average fidelity of $F_{\mathrm{avg}}=99.86(2)\%$. The extracted sensitive qubit (blue) fidelity is $F_{\mathrm{avg}}=99.78(3)\%$, but the data exhibit a poor fit to the decay model, as expected, with the fit line clearly failing to capture the loss of survival probability at larger gate set sizes.
This indicates that the fitted fidelity for the sensitive qubit is an over-estimate.
This behavior is consistent with deterministic line-synchronous detuning producing correlated errors during the randomized sequences, rather than acting as a purely stochastic error source. 

With compensation enabled the magnetic-field-insensitive qubit remains essentially unchanged at $F_{\mathrm{avg}}=99.84(2)\%$, while the performance of the magnetic-field-sensitive qubit improves substantially, as shown in Fig.~\ref{fig:rb_compensation}(b). 
The extracted average fidelity increases to $F_{\mathrm{avg}}=99.93(1)\%$. 
In contrast to the uncompensated case, the compensated sensitive-qubit data are well described by the randomized-benchmarking fit, indicating that the dominant non-Markovian line-synchronous contribution has been removed.
The sensitive qubit then performs comparably to, and in this data slightly better than, the magnetic-field-insensitive qubit.
This is expected because the sensitive qubit's transition is much stronger, so once the line-synchronous detuning is compensated, a large error source has been removed allowing the faster gates to outperform the slower, insensitive qubit thanks to lower exposure to residual decoherence during each operation.

These randomized benchmarking measurements show that the compensation protocol improves not only Ramsey-type characterization experiments, but also randomized gate sequences. 
The result is consistent with the interpretation that the line-synchronous signal acts as a coherent, repeatable, non-Markovian error source, and that referencing the control updates to the line trigger converts this error into a predictable correction.

\begin{figure*}[t] 
  \includegraphics[width=0.97\textwidth]{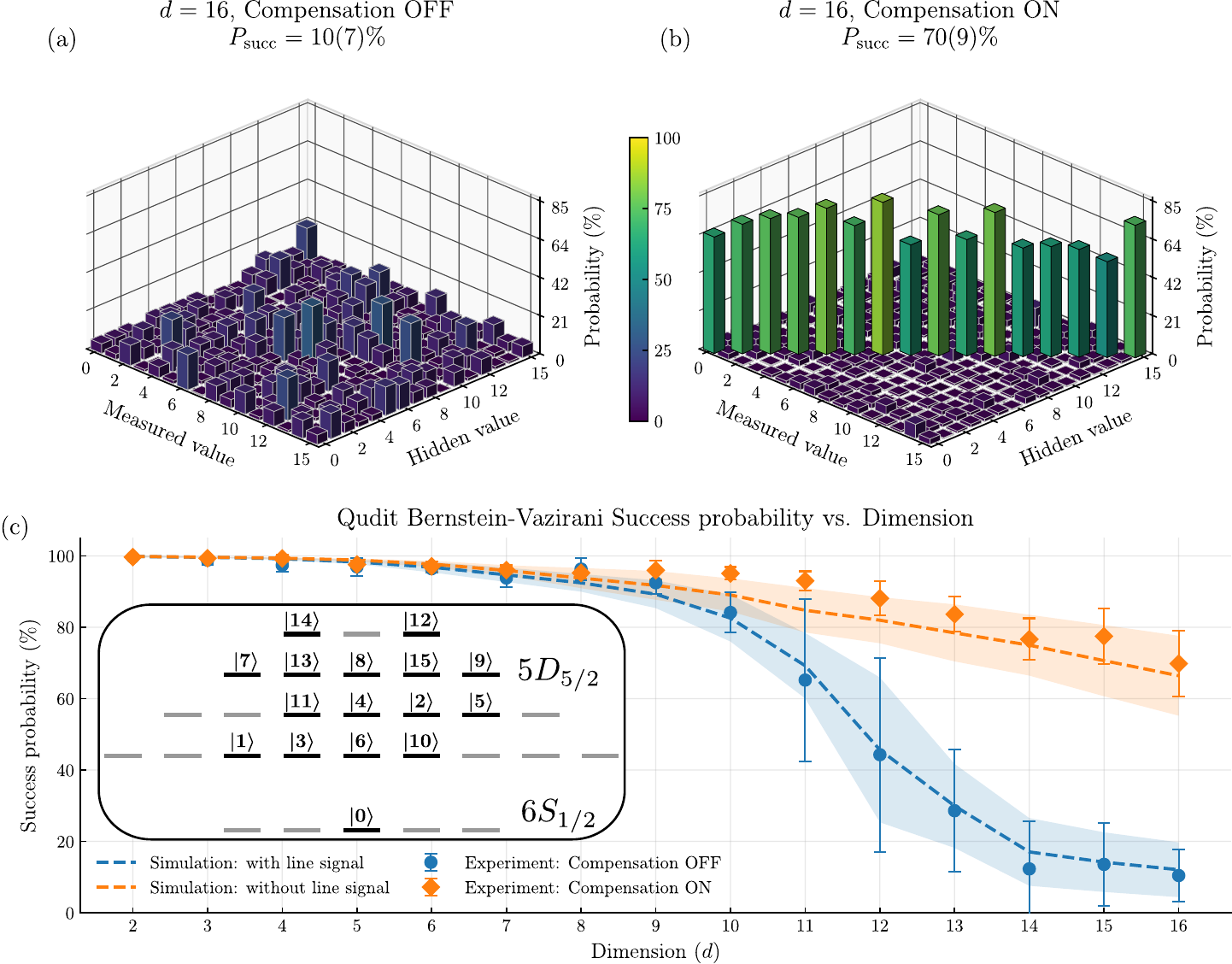}
    \caption{
    Qudit Bernstein--Vazirani algorithm performance with and without line-synchronous compensation. 
    (a) Measured output distribution for a $d=16$ qudit Bernstein--Vazirani experiment with AC line-synchronous compensation disabled. 
    The probability is broadly distributed over incorrect measured values, giving an average success probability of $P_{\mathrm{succ}}=10(7)\%$. 
    (b) Output distribution for the same $d=16$ experiment with AC line-synchronous compensation enabled. 
    The population is concentrated along the diagonal, where the measured value equals the hidden value, increasing the success probability to $P_{\mathrm{succ}}=70(9)\%$. 
    The color scale in panels (a) and (b) denotes the measured probability in percent. 
    (c) Success probability as a function of qudit dimension. 
    Blue circles show experimental results with compensation disabled, while orange diamonds show results with compensation enabled. 
    Dashed curves show simulations with the line signal included or removed, and shaded regions indicate the corresponding simulation spread, for simulation details see Methods~\ref{sec:qudit_BVA}.
    The inset shows the set of $^{137}\mathrm{Ba}^+$ states used for the qudit encoding up to $d=16$. 
    The success probabilities in the compensated data remain substantially higher at large dimension, demonstrating that pulse-by-pulse detuning and phase updates improve algorithmic performance in multilevel qudit circuits.
    }
    \label{fig:bva_compensation}
\end{figure*}

\subsection{Qudit algorithm performance}

Finally, we apply the qudit extension of the compensation protocol described in Methods~\ref{sec:extension_to_qudits} to test whether line-synchronous compensation improves the performance of a complete multilevel quantum algorithm.
Quantum algorithms are built from coherent sequences of gates, so their performance depends not only on whether individual pulses are resonant, but also on whether phase relationships are preserved throughout the full circuit. 
This makes an algorithmic experiment a useful end-to-end test of the compensation protocol.
We use the Bernstein--Vazirani algorithm implemented on a single $d$-level qudit~\cite{wang20_qudit_high_dimen_quant_comput}.
The Bernstein--Vazirani problem is an oracle-identification task: an unknown hidden value $h$ is encoded in the phase response of an oracle, and the goal is to determine $h$ from the final measurement. 
In the ideal quantum circuit, the hidden value can be recovered deterministically after one oracle query. 
Here, this makes the measured success probability a direct probe of whether coherent phase information is preserved across the full qudit operation.
The circuit consists of a quantum Fourier transform (QFT), followed by a hidden-value oracle, followed by an inverse quantum Fourier transform (QFT$^\dagger$) and measurement in the computational basis. 
Starting from the initial state $|0\rangle$, the first QFT prepares an equal superposition over all computational-basis states,
\begin{equation}
|0\rangle \rightarrow \frac{1}{\sqrt{d}}\sum_{x=0}^{d-1}|x\rangle .
\end{equation}
The oracle encodes a hidden value $h$ by applying a phase that depends on the basis-state index,
\begin{equation}
O_h |x\rangle = e^{2\pi i h x/d}|x\rangle.
\end{equation}
The inverse quantum Fourier transform (QFT$^\dagger$) then evolves the state to the computational basis state $|h\rangle$, an ideal measurement returns the hidden value $h$ with unit probability.

In this implementation, the test is especially sensitive to line-synchronous errors because the circuit requires coherent control across a multilevel magnetic field sensitive Hilbert space. 
As the qudit dimension increases, the circuit involves more basis states and more driven transitions, each with its own transition frequency, magnetic-field sensitivity, and accumulated phase. 
Line-synchronous perturbations can therefore reduce the algorithm success probability through both instantaneous detuning errors during pulses and phase errors on the pulses. 

We quantify performance using the success probability
\begin{equation}
P_{\mathrm{succ}}
=
\frac{1}{d}
\sum_{h=0}^{d-1}
P(\mathrm{measured}=h \mid \mathrm{hidden}=h),
\end{equation}
where the average is taken over all hidden values. 

Fig.~\ref{fig:bva_compensation} compares the measured Bernstein--Vazirani output distributions with and without line-synchronous compensation. 
The inset in Fig.~\ref{fig:bva_compensation}(c) shows the state selection used to encode the qudit basis up to $d=16$.
Each qudit of dimension $d$ is composed of the states $\{ |0\rangle, \cdots, |d-1\rangle \}$.
The computational state $|0\rangle$ is chosen from the $6S_{1/2}$ manifold, while the remaining basis states are selected from the $5D_{5/2}$ manifold, allowing the accessible Hilbert space dimension to be increased while retaining direct optical control over the required transitions. 
Without compensation, the output distribution is strongly distorted for large Hilbert space dimension. 
For example, at $d=16$, population is spread over many incorrect measured values, and yields a success probability of only $P_{\mathrm{succ}}=10(7)\%$ (within error of random chance). 
With compensation enabled, the population concentrates along the diagonal corresponding to the correct hidden value, increasing the success probability to $P_{\mathrm{succ}}=70(9)\%$. 
The improvement is also visible across dimension. 
At small $d$, both compensated and uncompensated circuits perform well. 
As $d$ increases, the uncompensated success probability decreases rapidly, consistent with the accumulation of line-synchronous detuning and phase errors across the multilevel control sequence.
With compensation applied, the success probability remains high across accessible dimensions, closely following the simulated behavior (dashed line) when line-synchronous effects are removed.

These results demonstrate that the compensation protocol extends beyond isolated two-level measurements and randomized gate sequences. 
Because the correction is applied pulse by pulse using the magnetic-field sensitivity of each driven transition, the same measured line-synchronous waveform can be used across a multi-tone, multilevel qudit operation. 
The resulting improvement in Bernstein--Vazirani success probability shows that software-defined line-synchronous compensation can directly enhance algorithmic performance.

\section{Discussion}

We have demonstrated a software-based approach to mitigating trigger-synchronous effects in a laboratory setting. 
This approach is cheap, in that no additional equipment or system re-design and re-configuration is needed for its implementation. 
It is easy to realize, just requiring a method to determine transition frequencies, which is a fundamental prerequisite to controlling a quantum system, frequency and phase control over the control pulses applied, and a means of triggering experiments 
to the same phase of the perturbation signal.
The approach is widely applicable, in that any qubits experiencing frequency or phase decoherence (i.e. most platform types) can benefit from it.
And finally, it is compatible with all other techniques aimed at reducing noise in quantum systems, due to its unobstructive, software-based nature.

All experiments shown in this work used experimental shots triggered to the same position on the AC mains line signal.
Such trigger-referencing introduces dead time in the processor duty-cycle, which can be especially tedious when triggers are separated by longer durations, such as the $\sim 1$ second periods of pulse-tube coolers  \cite{kono24_mechan_induc_correl_error_super, kalra16_vibrat_induc_elect_noise_cryog, kellermann24_vibrat_decoup_system_tes_operat, haan14_atomic_resol_scann_tunnel_micros}. 
By continuously tracking the trigger-synchronous signal in the control framework, it would be possible to recover this dead time and implement this technique regardless of where in the line signal each experimental shot begins.

The qudit demonstrations above showed that multiple transitions can be tracked and compensated simultaneously, which motivates applying this method to a register of qubits or qudits. 
Future work may be needed, though, to characterize the homogeneity of these types of coherent perturbations across a spatially distributed register, which will also inform the calibration overhead for multi-qubit or -qudit register.

Finally, the overall effectiveness of this technique is limited by two factors: the reproducibility of the trigger-synchronous signal, and the precision with which the signal can be characterized. 
Other sources of noise in our system limited this precision, since overall coherence times impact the precision of frequency calibrations. 
Implementing this technique along with other noise suppression methods, like magnetic-field shielding to suppress A/C mains fluctuations for instance, makes it possible to characterize trigger-synchronous signals more precisely, thereby enabling even better compensation and overall performance.

\appendix*
\renewcommand{\appendixname}{}
\section*{METHODS}

\subsection{Experimental setup}
\label{sec:experimental_setup}

We employ a two-step, isotope selective ablation loading scheme using 554~nm and 389~nm excitation~\cite{white22_isotop_selec_laser_ablat_ion, greenberg24_trapp_ba_with_seven_fold} to confine single ions in our custom-built four-rod linear Paul trap, with secular frequencies of (1.3, 1.5, 0.2)~MHz.
Further technical information about the ion trapping apparatus is described in \cite{low25_contr_readout_trapp_ion_qudit}.
In this work, the laser intensities are 13~mW/mm$^2$ for 493~nm, 113~mW/mm$^2$ for 650~nm, and 1.68~mW/mm$^2$ for 614~nm.
The magnetic field for all experiments is $\sim 4.209$G.
The magnetic field is supplied by an array of nickel-plated neodymium permanent magnets (Nd$_2$Fe$_{14}$B) held in 3D printed mounts fixed directly to the trap vacuum chamber.
Each experimental shot begins with Doppler cooling, followed by narrow-band optical-pumping based initialisation \cite{an22_high_fidel_state_prepar_measur, low25_quant_logic_operat_algor_singl}, control pulses, and ends with fluorescence-based readout \cite{low25_quant_logic_operat_algor_singl}.

The line trigger used for synchronization is derived from the laboratory mains signal through an electrically isolated low-voltage pickup circuit. 
A plug-in Class~2 transformer rated for $120~\mathrm{VAC}$, $60~\mathrm{Hz}$, $4~\mathrm{W}$ input and $5~\mathrm{VAC}$, $500~\mathrm{mA}$ output provides a low-voltage copy of the mains waveform. 
This signal is passed through an additional isolation transformer (Triad Magnetics N-51X) and then coupled through a ZFBT-6GW+ bias tee. 
The low-frequency output of the bias tee is sent to a TLV3501 comparator, which converts the sinusoidal line signal into a digital square-wave trigger. 
This trigger defines the line-referenced time origin used for both calibration and compensation measurements.

System calibration proceeds as follows:
Frequencies used to drive the quadrupole $6S_{1/2} \leftrightarrow 5D_{5/2}$ transitions are calibrated by measuring two reference transitions which enable calibration of magnetic field drifts as well as slow laser cavity drifts (see \cite{low25_quant_logic_operat_algor_singl} for details).
Frequency calibration is typically run every 2-3 minutes during data acquisition.
Transition strengths are calibrated by measuring the Rabi frequencies of five reference transitions, each with a unique $\Delta m \in \{-2, -1, 0, 1, 2\}$.
These serve as a reference for every other transition with the matching $\Delta m$ value, which are calculated from theory and scaled up/down according to the outputs of the five reference measurement.

\subsection{Measurement of the line-synchronous waveform}
\label{sec:measure_line_sig}

To measure the instantaneous line-synchronous detuning, each experimental shot is synchronized to the line trigger and followed by a variable delay $T_{\mathrm{LT}}$, as shown in Fig.~\ref{fig:detuning_phase_compensation}(a). 
After this delay, we apply a Ramsey sequence with fixed free-evolution time $\tau=100~\mu\mathrm{s}$ and measure the final population for two analyzer phases, $\phi=\pi/2$ and $\phi=3\pi/2$. 
Rather than converting the measured Ramsey phase to a detuning using the ideal short-pulse approximation, we fit the data to a numerical model of the full Ramsey sequence. 
For each candidate detuning, the model computes the driven evolution during the finite-duration $\pi/2$ pulses and the free evolution between them, and the detuning is extracted by minimizing the difference between the simulated and measured populations~\cite{low25_quant_logic_operat_algor_singl, tathed25_actual_high_dimens_qudit}.

To separate the line-synchronous perturbation from slow drifts, we measure two optical transitions at both the line-trigger reference time and at the delayed time $T_{\mathrm{LT}}$. 
The magnetic-field-sensitive transition is 
$|{6S_{1/2},F=2,m_F=0}\rangle \leftrightarrow |{5D_{5/2},F=4,m_F=-2}\rangle$, 
while the comparatively magnetic-field-insensitive transition is 
$|{6S_{1/2},F=2,m_F=0}\rangle \leftrightarrow |{5D_{5/2},F=2,m_F=0}\rangle$. 
Measuring both transitions allows slow ambient magnetic-field drift and laser-cavity drift to be tracked separately from the trigger-referenced line-synchronous signal. 
Repeating the measurement as a function of $T_{\mathrm{LT}}$ reconstructs the line-synchronous transition-frequency shift $\delta\omega(T_{\mathrm{LT}})$.

Because the magnetic-field sensitivity $\kappa=\partial\omega/\partial B$ of the probed transition is known from the calculated level structure~\cite{low25_contr_readout_trapp_ion_qudit, tathed25_actual_high_dimens_qudit}, the measured transition-frequency shift is converted into an equivalent magnetic-field perturbation,
\begin{equation}
\Delta B_{\mathrm{AC}}(t)
=
\frac{\delta\omega(t)}{2\pi\kappa}.
\end{equation}
This gives an experimentally calibrated magnetic-field waveform in the line-trigger-referenced time coordinate.

The line-synchronous waveform is sampled at 90 distinct trigger delays. 
The converted magnetic-field waveform is fit to a harmonic model restricted to integer multiples of the $60~\mathrm{Hz}$ line frequency,
\begin{equation}
\Delta B_{\mathrm{AC}}(t)
=
B_0+
\sum_{n=1}^{K}
A_n
\cos\!\left(2\pi n 60 t+\phi_n\right),
\end{equation}
where $B_0$ is a constant offset and each harmonic has an amplitude $A_n$ and phase $\phi_n$. 
We use the first ten harmonics of the $60~\mathrm{Hz}$ line frequency to fit the measured waveform.
This choice captures the visible non-sinusoidal structure of the line-synchronous signal while retaining substantially more sampled delays than fitted coefficients.
For 90 distinct trigger delays, the $K=10$ model contains $2K+1=21$ fit coefficients, leaving 69 degrees of freedom.
Fig.~\ref{fig:compensation_schematic}(e) shows the resulting harmonic fits to measurements taken on different dates, demonstrating the reproducibility of the line-synchronous magnetic-field waveform in the trigger-referenced time coordinate.

The resulting calibrated magnetic-field waveform can then be scaled by the appropriate magnetic-field sensitivity to predict the line-synchronous detuning of other transitions or between different states.

\subsection{Rotating-frame derivation of the compensation condition}
\label{sec:compensation_derivation}

We derive the compensation condition for a driven two-level transition whose resonance frequency is shifted by a deterministic line-synchronous perturbation. 
In a frame referenced to the line trigger, this perturbation is described by a time-dependent adiabatic angular-frequency shift $\delta_{\mathrm{AC}}(t)$. 
The goal is to show that a programmed phase ramp of the control oscillator is equivalent to an instantaneous frequency correction, and that the accumulated phase error is given by the time integral of the detuning.

Let $\omega_0$ denote the reference transition frequency. 
The applied control field is taken to be proportional to
\begin{equation}
\Omega(t)\cos\!\big(\theta_{\mathrm{L}}(t)\big),
\qquad
\theta_{\mathrm{L}}(t)=\omega_{\mathrm{L}}t+\phi_{\mathrm{L}}(t),
\end{equation}
where $\Omega(t)$ is the drive amplitude, $\omega_{\mathrm{L}}$ is the nominal oscillator frequency, and $\phi_{\mathrm{L}}(t)$ is a programmable phase. 
In the presence of the line-synchronous perturbation, the laboratory-frame Hamiltonian is
\begin{equation}
H(t)
=
\frac{\hbar}{2}\big[\omega_0+\delta_{\mathrm{AC}}(t)\big]\sigma_z
+
\hbar\Omega(t)\cos\!\big(\theta_{\mathrm{LO}}(t)\big)\sigma_x .
\end{equation}

To determine how the programmed oscillator phase enters the dynamics, we transform into the frame rotating with the instantaneous oscillator phase,
\begin{equation}
U(t)
=
\exp\!\left[-\frac{i}{2}\theta_{\mathrm{LO}}(t)\sigma_z\right].
\end{equation}
The rotating-frame Hamiltonian is
\begin{equation}
H_{\mathrm{rot}}(t)
=
U^\dagger(t)H(t)U(t)
-
i\hbar U^\dagger(t)\dot U(t).
\end{equation}
Since $U(t)$ is generated by $\sigma_z$, the diagonal qubit term is unchanged by the unitary transformation, while the frame-transformation term contributes
\begin{equation}
-i\hbar U^\dagger(t)\dot U(t)
=
-\frac{\hbar}{2}\dot\theta_{\mathrm{LO}}(t)\sigma_z .
\end{equation}
Thus the diagonal part of the rotating-frame Hamiltonian is
\begin{equation}
\frac{\hbar}{2}
\left[
\omega_0+\delta_{\mathrm{AC}}(t)-\dot\theta_{\mathrm{LO}}(t)
\right]\sigma_z .
\end{equation}

The drive term also transforms under $U(t)$. 
It separates into a slowly varying co-rotating component and a rapidly oscillating counter-rotating component. 
After applying the rotating-wave approximation, the Hamiltonian becomes
\begin{equation}
H_{\mathrm{rot}}(t)
\approx
\frac{\hbar}{2}\Delta_{\mathrm{eff}}(t)\sigma_z
+
\frac{\hbar}{2}\Omega(t)\sigma_x ,
\end{equation}
where
\begin{equation}
\Delta_{\mathrm{eff}}(t)
=
\omega_0+\delta_{\mathrm{AC}}(t)-\dot\theta_{\mathrm{LO}}(t).
\end{equation}
Using $\dot\theta_{\mathrm{LO}}(t)=\omega_{\mathrm{LO}}+\dot\phi_{\mathrm{LO}}(t)$ gives
\begin{equation}
\Delta_{\mathrm{eff}}(t)
=
\omega_0+\delta_{\mathrm{AC}}(t)
-
\omega_{\mathrm{LO}}
-
\dot\phi_{\mathrm{LO}}(t).
\label{eq:effective_detuning}
\end{equation}

Eq.~\eqref{eq:effective_detuning} shows that a time-dependent programmed phase is equivalent to an instantaneous frequency shift of the control oscillator. 
If the nominal oscillator is chosen to be resonant with the reference transition, $\omega_{\mathrm{LO}}=\omega_0$, then the line-synchronous detuning is cancelled by choosing
\begin{equation}
\dot\phi_{\mathrm{LO}}(t)
=
\delta_{\mathrm{AC}}(t).
\end{equation}
Equivalently, the programmed phase must satisfy
\begin{equation}
\phi_{\mathrm{LO}}(t)
=
\phi_{\mathrm{LO}}(t_0)
+
\int_{t_0}^{t}
\delta_{\mathrm{AC}}(t')\,dt' .
\label{eq:phase_accumulator}
\end{equation}
Here $t_0$ denotes the time at which the phase accumulator is initialized; in the experiments shown in this work we take $t_0=0$, the line-trigger-referenced start of the sequence.
Thus, compensation can be implemented either as a continuous frequency correction or as an accumulated phase update; these are equivalent descriptions of the same control operation.

In a frame rotating only at the fixed reference frequency, the same perturbation appears as a time-dependent detuning. 
In the absence of compensation, this detuning produces a deterministic $z$-rotation. 
Taking the start of the sequence, $t=0$, as the phase reference, the accumulated line-synchronous phase is
\begin{equation}
\Phi_{\mathrm{AC}}(t)
=
\int_0^t
\delta_{\mathrm{AC}}(t')\,dt'.
\end{equation}
This is the phase that must be tracked by the control system in order to keep later pulses phase coherent with the globally referenced oscillator.

For piecewise pulse sequences, Eq.~\eqref{eq:phase_accumulator} is implemented as a discrete phase accumulator evaluated at the pulse times. 

\subsection{Extension of compensation framework to qudits}
\label{sec:extension_to_qudits}

For a multilevel system, it is useful to view the line-synchronous perturbation as producing state-dependent phase accumulation. 
Consider a qudit state
\begin{equation}
|\psi(0)\rangle=\sum_{i=0}^{d-1} c_i |i\rangle .
\end{equation}
If the energy of state $|i\rangle$ is shifted by $\delta E_i(t)$, then in the absence of compensation the state evolves as
\begin{equation}
|\psi(t)\rangle
=
\sum_{i=0}^{d-1}
c_i
\exp\!\left[
-\frac{i}{\hbar}
\int_0^t \delta E_i(t')\,dt'
\right]
|i\rangle .
\end{equation}
Thus, a line-synchronous perturbation produces relative phases between qudit basis states.

For a pulse coupling two states $|m\rangle$ and $|n\rangle$, the relevant quantity is the differential phase accumulated between those states. 
This is determined by the transition-specific detuning
\begin{equation}
\delta\omega_{mn}(t)
=
\frac{\delta E_n(t)-\delta E_m(t)}{\hbar}.
\end{equation}
Therefore, the qubit compensation rule applies directly to each driven qudit transition by using the detuning waveform associated with that transition.

When the perturbation is dominated by magnetic-field noise and the field variation is small enough that the transition frequency depends linearly on magnetic field, this transition-specific detuning can be written as
\begin{equation}
\delta\omega_{mn}(t)=2\pi\kappa_{mn}\Delta B(t),
\end{equation}
where $\Delta B(t)$ is the measured line-synchronous magnetic-field deviation and $\kappa_{mn}$ is the magnetic-field sensitivity of the $|m\rangle\leftrightarrow |n\rangle$ transition. 
Thus, all driven transitions use the same measured field waveform, but each is scaled by its own transition sensitivity.

For pulse $j$ driving transition $|m_j\rangle\leftrightarrow |n_j\rangle$ at time $t_j$, the programmed oscillator frequency is
\begin{equation}
\omega_{L,j}^{(\mathrm{prog})}
=
\omega_{m_j n_j}^{(0)}
+
2\pi\kappa_{m_j n_j}\Delta B(t_j),
\end{equation}
where $\omega_{m_j n_j}^{(0)}$ is the reference transition frequency at the global sequence start. 
The corresponding programmed phase is
\begin{equation}
\phi_j^{(\mathrm{prog})}
=
\phi_j^{(\mathrm{ideal})}
+
2\pi\kappa_{m_j n_j}
\left[
\int_0^{t_j}\Delta B(t')\,dt'
-
t_j\Delta B(t_j)
\right].
\end{equation}
with $\phi_j^{(\mathrm{ideal})}$ set by the desired qudit gate decomposition.

This pulse-by-pulse prescription is especially useful for qudit circuits because a single algorithm may involve many transitions with different magnetic-field sensitivities. 
Line-synchronous noise can therefore produce both transition-dependent detuning errors during pulses and relative phase errors between computational basis states. 
By applying the correction separately for each addressed transition, using the same trigger-referenced field waveform, the control sequence remains phase-consistent across the multilevel Hilbert space. 
No additional line-signal measurement is required when changing the qudit dimension or circuit, provided that the transition sensitivities are known.

\subsection{Quantifying suppression of line-synchronous effects}
\label{sec:compensation_quantification}
\subsubsection{Matched-filter projection of the calibrated AC waveform}
\label{sec:matched_filter_projection}

Matched filtering is a standard signal-processing technique for detecting or estimating the amplitude of a known waveform in noisy data \cite{turin60_introd_to_match_filter}.
It is commonly used when the expected time dependence of a signal is known in advance, and the goal is to determine how strongly that waveform is present in a measured record \cite{allen12_findc, gibbons06_detec_low_magnit_seism_event}. 
Here, the independently calibrated AC line signal provides the known waveform. 
Rather than asking only whether the measured detuning or phase has a large variance, we project the measured data onto the calibrated AC waveform and use the fitted projection amplitude to quantify how much of the deterministic AC contribution remains after compensation.

To quantify how much of the calibrated AC line signal remains after compensation, we project the measured residual signal onto the independently measured AC waveform. 
For the detuning-compensation experiment, the calibrated magnetic-field waveform $\Delta B_{\mathrm{AC}}(t)$ is converted into the expected transition-frequency shift,
\begin{equation}
f_{\mathrm{AC}}(t)
=
\kappa\Delta B_{\mathrm{AC}}(t),
\end{equation}
where $\kappa$ is the magnetic-field sensitivity of the transition under study. 
The measured detuning is then fit to
\begin{equation}
\delta\omega_{\mathrm{meas}}(t)
=
b
+
a_{\mathrm{AC}} f_{\mathrm{AC}}(t)
+
\epsilon_i .
\label{eq:detuning_matched_filter}
\end{equation}
Here $b$ accounts for a static detuning offset, while $a_{\mathrm{AC}}$ gives the amplitude of the calibrated AC line-synchronous waveform remaining in the measured data. 
In the absence of compensation, $a_{\mathrm{AC}}\simeq 1$ indicates that the measured detuning follows the calibrated AC waveform. 
With ideal compensation, the deterministic AC contribution is removed and $a_{\mathrm{AC}}\simeq 0$.

For the phase-compensation measurement, the corresponding AC phase waveform is obtained from the accumulated detuning,
\begin{equation}
\Phi_{\mathrm{AC}}(t)
=
\int_0^t \delta\omega_{\mathrm{AC}}(t')\,dt' .
\end{equation}
The measured phase is then fit to
\begin{equation}
\phi_{\mathrm{meas}}(t)
=
b
+
m t
+
a_{\mathrm{AC}}\Phi_{\mathrm{AC}}(t)
.
\label{eq:phase_matched_filter}
\end{equation}
Here $b$ accounts for a constant phase offset, $m$ accounts for any residual static detuning, and $a_{\mathrm{AC}}$ gives the amplitude of the calibrated AC phase waveform remaining in the measured data.

For both detuning and phase measurements, we define the matched-filter suppression factor as
\begin{equation}
\mathcal{S}_{\mathrm{MF}}
=
\frac{
|a_{\mathrm{AC}}^{\mathrm{off}}|
}{
|a_{\mathrm{AC}}^{\mathrm{on}}|
},
\end{equation}
where $a_{\mathrm{AC}}^{\mathrm{off}}$ and $a_{\mathrm{AC}}^{\mathrm{on}}$ are the fitted AC amplitudes without and with compensation, respectively.

\subsubsection{Suppression of harmonic AC content}
\label{sec:harmonic_suppression}

As a complementary measure to the matched-filter projection, we quantify how much periodic structure remains at the line frequency and its harmonics. 
This metric does not assume that the residual signal has the same shape as the calibrated AC waveform; instead, it measures the total fitted amplitude at harmonics of the line frequency.

For detuning data, the measured signal is fit to
\begin{equation}
\delta\omega_{\mathrm{meas}}(t)
=
c_0
+
\sum_{n=1}^{K}
A_n
\cos\!\left(2\pi n 60t+\phi_n\right).
\label{eq:detuning_harmonic_metric}
\end{equation}
For phase data, we include an additional linear term to account for any residual static detuning,
\begin{equation}
\phi_{\text{meas}}(t)
=
c_0
+
c_1t
+
\sum_{n=1}^{K}
A_n
\cos\!\left(2\pi n 60t+\phi_n\right).
\label{eq:phase_harmonic_metric}
\end{equation}
The constant offset $c_0$ and linear slope $c_1$ are used only to remove non-oscillatory phase offsets before extracting the harmonic amplitudes. 
The harmonic content is quantified only from the fitted amplitudes $A_n$.

The total harmonic AC amplitude is defined as
\begin{equation}
A_{\mathrm{AC}}
=
\sqrt{
\sum_{n=1}^{K} A_n^2
}.
\end{equation}
We then define the harmonic-amplitude suppression as
\begin{equation}
\mathcal{S}_{A}
=
\frac{
A_{\mathrm{AC}}^{\mathrm{off}}
}{
A_{\mathrm{AC}}^{\mathrm{on}}
},
\end{equation}
where $A_{\mathrm{AC}}^{\mathrm{off}}$ and $A_{\mathrm{AC}}^{\mathrm{on}}$ are the fitted harmonic amplitudes without and with compensation, respectively. 
The matched-filter projection measures suppression of the independently calibrated AC waveform, while this harmonic metric measures suppression of residual line-frequency structure more generally.

\subsection{Unitary gate decomposition algorithm}
\label{sec:gate_decomposition}

All unitary gates applied in the randomized benchmarking experiments and Bernstein--Vazirani algorithms are implemented via sequences of quadrupole transitions coupling the $6S_{1/2} \leftrightarrow 5D_{5/2}$ manifolds, as mentioned throughout the text. 
In order to implement a target unitary gate $U_t$ in this context, we use an efficient pulse sequence finding algorithm which is based on QR decomposition and is ``transition-aware'', adapted from \cite{drozhzhin25_trans_aware_decom_singl_qudit_gates}.

To begin, we note explicitly the native operations available to our platform, given the state sets used as shown in the inset of Fig.~\ref{fig:bva_compensation}c. 
These are (1) a phase shift gate $\mathrm{P}^j$ on basis state levels (which can be implemented virtually on our platform) and (2) two-level transitions $\mathrm{R}^{kl}$(Givens rotations with complex phases), expressible as

\begin{align}
    \mathrm{P}^j(\theta) &= \exp\left\{ i\theta \left|j\right\rangle \left\langle j\right| \right\} \\
    \mathrm{R}^{kl}(\theta, \phi) &= \exp\left\{ -i\frac{\theta}{2} \sigma^{kl}_\phi \right\}
\end{align}
where
\begin{equation}
    \sigma^{kl}_\phi = e^{+i\phi} \left|k\right\rangle \left\langle l\right| + e^{-i\phi} \left|l\right\rangle \left\langle k\right|.
\end{equation}

The decomposition scheme proceeds in two steps. First, a recipe of which transitions to implement (and in what order), independent of any target unitary gate, is determined based on the connection graph.
In the star topology this eliminates off-diagonal elements row by row, starting from the bottom up.

\vspace{1em}
\noindent\fbox{%
\begin{minipage}{\linewidth}
\begin{center}
\label{alg:transition-aware-unitary-decomposition}
  \textbf{Algorithm 1} \\
  Transition-aware star decomposition
\end{center}
\hrule
\vspace{0.5em}

\textbf{Require:} $U \in \mathrm{U}(d)$, tolerance $\varepsilon = 10^{-12}$

\begin{tabbing}
1234\=12\=12\=12\= \kill
\textbf{1:}  \> $V \leftarrow U$ \\
\textbf{2:}  \> $\mathcal{P} \leftarrow [\ ]$ \quad \textit{(pulse record)} \\
             \textbf{--- Scheme-driven column elimination ---} \\
\textbf{3:}  \> \textbf{for} $r \in \{d{-}1,\,\ldots,\,1\}$ \textbf{do} \\
             \>\> \textit{(off-hub rows; pivot $= 0$)} \\
\textbf{4:}  \>\> \textbf{for} $c \in \{1,\,\ldots,\,r{-}1\}$ \textbf{do} \\
\textbf{5:}  \>\>\> $x \leftarrow V_{0,\,r}$;\ $y \leftarrow V_{c,\,r}$ \\
\textbf{6:}  \>\>\> \textbf{if} $\sqrt{|x|^2+|y|^2} > \varepsilon$ \textbf{then} \\
\textbf{7:}  \>\>\>\> $\theta \leftarrow 2\arctan2(|y|,|x|)$;\ \\
             \>\>\>\> $\phi \leftarrow \angle x - \angle y$ \\
\textbf{8:}  \>\>\>\> $G \leftarrow R_{0,c}(\theta,\phi)$;\ \\
             \>\>\>\> $V \leftarrow G^\dagger V$;\ \\
             \>\>\>\> $\mathcal{P}.\mathrm{append}\!\left((0,\,c),\,\theta,\,\phi\right)$ \\
             \textbf{--- Zero hub entry $V_{0,r}$; pivot $= V_{r,r}$ ---} \\
\textbf{9:}  \>\> $x \leftarrow V_{r,\,r}$;\ $y \leftarrow V_{0,\,r}$ \\
\textbf{10:} \>\> \textbf{if} $\sqrt{|x|^2+|y|^2} > \varepsilon$ \textbf{then} \\
\textbf{11:} \>\>\> $\theta \leftarrow 2\arctan2(|y|,|x|)$;\
             $\phi \leftarrow \angle x - \angle y$ \\
\textbf{12:} \>\>\> $G \leftarrow R_{r,0}(\theta,\phi)$;\ \\
             \>\>\> $V \leftarrow G^\dagger V$;\ \\
             \>\>\> $\mathcal{P}.\mathrm{append}\!\left((r,\,0),\,\theta,\,\phi\right)$ \\
             \textbf{--- Fold phases into pulses ---} \\
\textbf{13:} \> $\mathrm{P_{\text{tot}}} \leftarrow \bigl(\angle\, V_{kk}\bigr)_{k=0}^{d-1}$
             \quad \textit{(phases on diag. of $V$)} \\
\textbf{14:} \> \textbf{for} $\bigl((i,j),\,\theta,\,\phi\bigr)$ in $\mathcal{P}$ \textbf{do} \\
\textbf{15:} \>\> \textbf{if} $i > j$ \textbf{then} \quad
             $i,j \leftarrow j,i$;\
             $\theta,\phi \leftarrow {-\theta},{-\phi}$ \\
\textbf{16:} \>\> $\phi \leftarrow \phi + (\lambda_j - \lambda_i)$ \\
\textbf{17:} \> \textbf{return} $\mathcal{P},\ \mathrm{P_{\text{tot}}}$
\end{tabbing}
\end{minipage}%
}
\vspace{1em}

Considering a $d$-level system with transitions $\mathrm{R}^{0n}$, with $0 < n < d$, we proceed by first eliminating all row elements in a chosen, target $\mathcal{U}$'s $d-1^{th}$, working up to applying $\mathrm{R}^{0,d-1}$ to eliminate the zeroth entry too. 
The elimination of each row, starting at row $d$, requires $d-1$ $\mathrm{R}$ gates, and fixing the final phase of the remaining diagonal entry requires a $\mathrm{P}$ gate.
As a result, the total number of individual native operations for a unitary with $d^2$ entries is $N_\mathrm{R} = d(d-1)/2$ and $N_\mathrm{P} = d$.
Since the phase shift gates can be applied virtually, this yields a quadratic scaling in the number of native operations needed for a qudit of dimension $d$.
The full implementation is given explicitly in Algorithm 1.

There are several details to notice in this implementation.
Firstly, as each rotation is applied, individual entries in the target unitary will change as a result of pulling out each individual $\mathrm{R}^{k,l}$ operation. 
This is indicated by the variable superscripts assigned to matrix elements throughout the four-step process shown. The final $\mathrm{P}^{j}$ phase shift applied per row is interspersed with subsequent rotations to zero out entries in row $d-2$ up to row 1, but all these phase shifts commute with these subsequent rotations by construction given the order in which rows are cancelled here. 
As a result, the $d$ phase shift gates needed here can all be gathered together as a diagonal phase fixing matrix and applied virtually after all pulses, resulting in a total pulse decomposition scheme which looks like 

\begin{equation}
\begin{aligned}
  \label{eq:ququart-taqr-decomp}
  \mathcal{U}_{\text{target}} = 
  &\mathrm{P}^0 \circ \mathrm{P}^1 \circ \mathrm{R}^{0,1} \circ
    \mathrm{P}^2 \circ \mathrm{R}^{0,2} \circ \mathrm{R}^{1,0}
    \circ \mathrm{P}^3 \\
    &\hspace{3.2em} \circ \mathrm{R}^{0,3} \circ
    \mathrm{R}^{2,0} \circ \mathrm{R}^{1,0} \\
= &\mathrm{P_{\text{tot}}} \circ \mathrm{R}^{0,1} \circ
    \mathrm{R}^{0,2} \circ \mathrm{R}^{1,0}
    \circ \mathrm{R}^{0,3} \circ
    \mathrm{R}^{2,0} \circ \mathrm{R}^{1,0}.
\end{aligned}
\end{equation}
where $\mathrm{P_{\text{tot}}}$ is the diagonal matrix of phases fixing all rows in the unitary.
Finally, as indicated in the algorithm steps, it is critical that one set a convention for the indices on rotations (which is enforced in line 15 of the algorithm table for instance) such that implemented rotations like $\mathrm{R}^{0,j}$ and $\mathrm{R}^{j,0}$ have consistent values for $\theta$ and $\phi$ in the Givens rotation matrices.

\subsection{Randomized benchmarking methods}
\label{sec:rb_method}

\subsubsection{Generating Haar-random matrices}

Starting from a general input dimension $d$, we first generate a complex matrix given by
\begin{equation}
Z = \frac{X + iY}{\sqrt{2}},
\end{equation}
where $X$ and $Y$ are real $d \times d$ matrices with independent and identically distributed (i.i.d.) standard normal entries. Each entry of $Z$ is therefore an i.i.d. complex normal random variable such that the variance of these entries tends to 1 with increasing number of matrices $Z$ included
\cite{mezzadri06_how_to_gener_random_matric}. 

We then compute the $QR$ decomposition where $Q$ is unitary and $R$ is upper triangular. 
However, the raw $Q$ returned by a numerical QR decomposition is not yet guaranteed to follow the Haar distribution because the diagonal entries of $R$ may carry arbitrary complex phases (the results of a QR decomposition are not generally unique)
\cite{edelman05_random_matrix_theor}.
To fix this problem, we impose another common convention, namely that the diagonal entries of $R$ are real. That is, the diagonal entries of $R$ are extracted and normalized:
\begin{equation}
\lambda_j = \frac{R_{jj}}{|R_{jj}|}.
\end{equation}

These phase factors are then collected into a diagonal matrix
\begin{equation}
\Lambda = \mathrm{diag}(\lambda_1,\ldots,\lambda_d),
\end{equation}

which is applied to $Q$ and $R$ to yield the corrected unitary $U$ and upper right triangular matrix $R'$
\begin{align}
U &= Q\Lambda \\
R' &= \Lambda^* R.
\end{align}
This removes the arbitrary diagonal phase convention from the QR factorization and yields a matrix distributed according to the Haar measure on $U(d)$.

\subsubsection{Implementing qubit unitary gates}

The randomized benchmarking results shown for two different optical qubit encodings in this work are implemented using the unitary gate decomposition algorithm from Methods~\ref{sec:gate_decomposition}. 
In this method, for the qubit case where the primitive operation consists of a rotation about a unit
vector in the $xy$-plane of the Bloch sphere defined by some angle $\phi$, with a rotation angle set by $\theta$, we have the freedom to implement \textit{any} single unitary in a single pulse. 
Any qubit unitary will have four real, free parameters, and the variables $\theta$ and $\phi$ account for two of them. 
The other two are set by a phase shifts on the basis states after the applied rotation.

Put explicitly then, any single qubit unitary $\hat{U}_j$ in some set of unitaries of size $s$ is decomposed into
\begin{equation}
  \label{eq:qubit-TAQR-decomposition}
  \hat{U}_j = \Lambda_j(\xi_0,\xi_1) R_j(\theta, \phi),
\end{equation}
where $R_j(\theta, \phi)$ is the two-level transition with rotation angle $\theta$ and phase $\phi$, and $\Lambda_j$ is the diagonal phase matrix with phase shifts $\xi_0$ and $\xi_1$ that would need to be applied to each basis state \textit{but are not}. 
Instead, when decomposing a set of unitaries into a full pulse sequence, the leftover phase shifts from the $j^{\mathrm{th}}$ unitary are subsumed into the following  $j+1^{\mathrm{th}}$ unitary. 
Therefore, for any unitary $U_j$ for $j > 0$, we decompose the operator $U_j \Lambda_{j-1}$. 
In this manner, every diagonal phase shift matrix from each individual unitary gate is ``pulled through'' the gate sequence. 
This is also applied for the final, inverting unitary $\hat{U}_{\mathrm{tot}}^{-1}$ and the resulting measured survival probabilities are therefore unaffected by the one remaining phase shift matrix (which might be called $\Lambda_{\mathrm{tot}}^{-1}$). 
This decomposition scheme allows us to implement qubit unitaries (or unitaries in any dimension really) as efficiently as possible, exploiting the virtual-$Z$ gate type rotations that our free choice of phases in subsequent pulses give us.

\subsection{Qudit-native Bernstein--Vazirani algorithm}
\label{sec:qudit_BVA}

\subsubsection{State selection}
\label{sec:qudit_state_selection}

The encoding scheme for multi-level measurements follows the method outlined in \cite{low25_quant_logic_operat_algor_singl} in which the mutual coherence and connectivity of a set of states are the most important qualities to optimise over.
For this work, we confine the search for states to those in $5D_{5/2}$ with direct quadrupole transitions to the $|6S_{1/2},F=2,m_F=0\rangle$ state, thus enforcing a ``star-topology'' to our connection graph.

In order to evaluate a cost function to optimise over and find an appropriate set of states, we extract effective coherence times for each transition using measured magnetic field noise, as well as laser phase noise, and the magnetic field sensitivities of the energy levels themselves, to calculate the pairwise expected coherence due to three separate sources for a transition between two states $j$ and $k$: Gaussian laser phase/frequency noise $T_{L,G}^{j,k}$, Lorentzian laser phase/frequency noise $T_{L,L}^{j,k}$, and Gaussian magnetic field noise $T_{B,G}^{j,k}$.

For a set of states $\{S\}$, we then define the cost function as

\begin{equation}
  \mathcal{C}(\{S\}) = \frac{1}{l} \left(
    \sum\limits_{j<k} \frac{\tau_{\{S\}}^2}{(T_{L,G}^{j,k})^2}
  + \sum\limits_{j<k} \frac{\tau_{\{S\}}}{T_{L,L}^{j,k}}
  + \sum\limits_{j<k} \frac{\tau_{\{S\}}^2}{(T_{B,G}^{j,k})^2} \right)
\end{equation}
where $l$ is the number of states in the set (i.e. the dimension of the multi-level system used for that particular measurement), and $\tau_{\{S\}}$ is the sum over all the transition times between states in $\{S\}$.
With these constraints, and this cost function, we exhaustively search all possible sets of states of size $d$ and pick the set that minimises $\mathcal{C}$. Under the constraint that all state sets include the $|6S_{1/2},F=2,m_F=0\rangle$ state, the resulting sets form a nested hierarchy in which each optimal set of size $N$ is a superset of that with $N-1$ members, differing by just one new element. All of these found sets of size $N < 17$ are shown in the inset of Fig.~\ref{fig:bva_compensation}.

\subsubsection{Simulation details and noise parameters}
\begin{table}[t]
\centering
\caption{Noise and error sources included in the BVA simulation.}
\label{tab:bva_noise_sources}
\begin{tabular}{lccc}
\hline
Noise/error source & Notation & Profile & FWHM \\
\hline
Magnetic field noise & $\delta_B$ & Gaussian & $26~\mu\mathrm{G}$ \\
Laser-frequency noise & $\delta_L$ & Voigt & $295~\mathrm{Hz}$ \\
Calibration-frequency error & $\delta f$ & Lorentzian & $30~\mathrm{Hz}$ \\
Pulse-angle error & $\delta_p$ & Gaussian & $4.38\%$ \\
AC line signal & $\Delta B_{\mathrm{AC}}(t)$ & \multicolumn{2}{c}{Table~\ref{table:line_signal_fit_params}} \\
\hline
\label{table:noise_parameters}
\end{tabular}
\end{table}

\begin{table}[t]
\centering
\caption{
Fitted line-signal model used in the BVA simulation. 
The field is modeled as $\Delta B_{\text{AC}}(t)=B_0+\sum_n A_n \cos(2\pi f_n t+\phi_n)$. 
}
\begin{tabular}{ccc}
\hline
$f_n$ (Hz) & $A_n$ (mG) & $\phi_n$ (rad) \\
\hline
$B_0$ & $0.327(6)$ & -- \\
60  & $0.311(9)$ & $-2.35(3)$ \\
120 & $0.015(9)$ & $2.3(6)$ \\
180 & $0.083(9)$ & $2.5(1)$ \\
240 & $0.007(9)$ & $3(1)$ \\
300 & $0.033(9)$ & $-2.2(3)$ \\
360 & $0.007(9)$ & $-1(1)$ \\
420 & $0.011(9)$ & $1.7(8)$ \\
480 & $0.009(9)$ & $1.1(9)$ \\
540 & $0.014(9)$ & $0.9(6)$ \\
600 & $0.014(9)$ & $-0.6(6)$ \\
\hline
\label{table:line_signal_fit_params}
\end{tabular}
\end{table}

For the BVA simulations, we used a Monte Carlo propagation of the full compiled pulse sequence.
Each ideal algorithm unitary was first decomposed into a sequence of two-level rotations on the experimentally available star topology.
For each Monte Carlo shot, noise parameters are sampled from the distributions summarized in Table~\ref{table:noise_parameters}. 
The calibrated AC magnetic-field waveform $\Delta B_{\mathrm{AC}}(t)$ is evaluated using the harmonic fit parameters given in Table~\ref{table:line_signal_fit_params}.
For each shot, we first construct a diagonal global detuning Hamiltonian,
\begin{equation}
H_{\mathrm{global}}
=
\sum_{(a,b)\in \mathcal{E}}
\frac{2\pi\delta_{ab}}{2}
\left(
|a\rangle \langle a|
-
|b\rangle \langle b|
\right),
\end{equation}
where \(\mathcal{E}\) is the set of distinct addressed transitions in the compiled sequence. \(\delta_{ab}\) includes Laser-frequency noise and calibration error as common-mode detuning shifts, and magnetic-field noise is converted to detuning through the transition sensitivity
\begin{equation}
    \delta_{ab} = \delta_L + \delta_f + \kappa_{ab}\delta_B
\end{equation}
This diagonal Hamiltonian is present during every pulse in the shot. 
For pulse $j$ driving transition $|m_j\rangle\leftrightarrow |n_j\rangle$ at time $t_j$, we add the AC line-synchronous detuning for that addressed transition and the resonant drive term,
\begin{equation}
\begin{aligned}
H^{(j)}
&=
H_{\mathrm{global}}
+
\frac{2\pi\kappa_{mn} \Delta B_{\mathrm{AC}}(t_j)}{2}
\left(
|m\rangle\langle m|
-
|n\rangle\langle n|
\right)
\\
&\quad
+
\frac{(1+\delta_p)\Omega_{mn}}{2}
\left(
e^{i\phi_j^{\mathrm{tot}}}
|m\rangle\langle n|
+
e^{-i\phi_j^{\mathrm{tot}}}
|n\rangle\langle m|
\right).
\end{aligned}
\label{eq:pulse_hamiltonian}
\end{equation}
Here $\Omega_{mn}$ is the nominal Rabi-frequency of the $|m\rangle \leftrightarrow |n\rangle$ transition and $\delta_p$ is a dimensionless fractional pulse-angle error caused by beam angle and intensity fluctuations. The value of $\delta_p$ is sampled from the pulse-angle-error distribution listed in Table~\ref{tab:bva_noise_sources}.
The total phase $\phi_{j}^{\text{tot}}$ contains the ideal pulse phase and the phase accumulated from the calibrated AC magnetic-field signal, 
\begin{equation}
\phi_{j}^{\text{tot}}=
\phi_{j}^{\text{ideal}}
+
2\pi \kappa_{mn}
\int^{t_{j}}_{0} \Delta B_{\mathrm{AC}}(t')\,dt'.
\end{equation}
The corresponding pulse propagator unitary is
\begin{equation}
U_j
=
\exp\!\left(-iH^{(j)}\tau_j\right),
\end{equation}
where $\tau_j$ is the pulse duration.

The full simulated sequence is obtained by multiplying the pulse propagators in time order. For a sequence containing $N$ pulses, the final state is
\begin{equation}
|\psi_{f}\rangle
=
U_{N}U_{N-1}\cdots U_2U_1 |\psi_0\rangle ,
\end{equation}
where $|\psi_0\rangle =|0\rangle$ is the initial state used for the BVA circuit. The simulated output probabilities are then obtained from the final-state populations in the computational basis.
The reported simulated success probabilities are averages over the Monte Carlo samples. 

\subsection{Phase unwrapping}
\label{sec:phase_unwrapping}

Measured phases are first returned modulo $2\pi$ and are therefore unwrapped before extracting phase trends. 
For smoothly varying phase data, we use a trend-aware unwrapping procedure rather than choosing the branch closest only to the previous point. 
For the first few points, the phase branch is chosen by ordinary local continuity. 
After this initialization, the next phase value is predicted from a linear fit to the previous three unwrapped points. 
For each new wrapped phase $\phi_i$, candidate branches $\phi_i+2\pi k$ are generated over a finite range of integers $k$, and the branch closest to the predicted phase is selected. 
This procedure preserves continuity while reducing unwrap errors when the phase changes rapidly or when adjacent points are separated by more than $\pi$.

\section*{Data availability}
\label{sec:data_availability}

The data underlying this work have been deposited at the Zenodo database at \cite{gatathed26_line_signal_zenodo_dataset} with open access.
\vspace{0.5cm}

\section*{Code availability}
\label{sec:code_availability}

The code underlying this work has been deposited at the Zenodo database at \cite{gatathed26_line_signal_zenodo_dataset} with open access.
\vspace{0.5cm}
   
\section*{Acknowledgements}
\label{sec:acknowledgements}
We thank Pei Jiang Low for the initial suggestion that the line-synchronous perturbation could be treated as a deterministic signal and compensated in software.

This research was supported, in part, by the Natural Sciences and Engineering Research Council of Canada
(NSERC), RGPIN-2025-06439, and by an Ontario Early Researcher Award. C.S. is also supported by a Canada Research Chair.

\textbf{Author Contributions:} C.S. supervised the project. G.A.T., N.C.F.Z., and C.S. conceived experiments. G.A.T., C.J.C.E., and N.C.F.Z. conducted the experiments. All authors analysed the results and reviewed the manuscript.

\textbf{Competing Interests:} The authors declare no competing interests.

\bibliography{references}

\end{document}